\documentclass{PoS}

\title{Making Access to Astronomical Software More Efficient}

\ShortTitle{Access to Astronomical Software}

\author{\speaker{Preben Grosb{\o}l}\thanks{on behalf of the OPTICON
    Network on `Future Astronomical Software Environments'.}\\
        ESO, Karl-Schwarzschild Str.~2, D-85748 Garching, Germany\\
        E-mail: \email{pgrosbol@eso.org}}

\author{Douglas Tody\\
        NRAO, 1003 Lopezville Rd, Socorro NM, USA, \\
        E-mail: \email{dtody@nrao.edu}}

\abstract{Access to astronomical data through archives and VO is
  essential but does not solve all problems.  Availability of
  appropriate software for analyzing the data is often equally
  important for the efficiency with which a researcher can publish
  results.  A number of legacy systems (e.g. IRAF, MIDAS, Starlink,
  AIPS, Gipsy), as well as others now coming online are available
  but have very different user interfaces and may no longer be fully
  supported.  Users may need multiple systems or stand-alone packages
  to complete the full analysis which introduces significant overhead.

  The OPTICON Network on `Future Astronomical Software Environments'
  and the USVAO have discussed these issues and have outlined a general
  architectural concept that solves many of the current problems
  in accessing software packages.  It foresees a layered structure
  with clear separation of astronomical code and IT infrastructure.
  By relying on modern IT concepts for messaging and distributed
  execution, it provides full scalability from desktops to clusters
  of computers.  A generic parameter passing mechanism and common
  interfaces will offer easy access to a wide range of astronomical
  software, including legacy packages, through a single scripting
  language such as Python.

  A prototype based upon a proposed standard architecture is being
  developed as a proof-of-concept.  It will be followed by definition
  of standard interfaces as well as a reference implementation which
  can be evaluated by the user community.  For the long-term success of
  such an environment, stable interface specifications and adoption
  by major astronomical institutions as well as a reasonable level
  of support for the infrastructure are mandatory.  Development and
  maintenance of astronomical packages would follow an open-source,
  Internet concept.  }

\FullConference{Accelerating the Rate of Astronomical Discovery, SpS5\\
		August 11-14, 2009\\
		Rio de Janeiro, Brazil}

\begin{document}

\section{Introduction}
The efficiency and speed with which one can obtain new astronomical results
depend upon a long chain of facilities and tools such as telescopes,
instruments, data archives, and software packages.  Although many discoveries
are made using new, more capable telescopes and instruments, one should not
underestimate the importance of easy access to both archival data and
state-of-the-art software.  The combination of multi-wavelength data (e.g.
obtained through data archives) and application of new algorithms for analysis
of data can lead to new understanding of astronomical phenomena.

In this paper, we focus on how access to and sharing of astronomical
software can be made easier and thereby increase the speed with which
new results are obtained. The prime concern is the efficiency of
turning raw data into astronomically relevant information. First an
overview of the current situation is given, followed by considerations
on possible ways to improve it.  An architectural concept for a future
astronomical software environment is then presented as discussed
by OPTICON and the USVAO (formerly NVO).  Finally, steps needed
to provide the astronomical community with an efficient software
environment for future data challenges are reviewed.

\section{Current state of affairs}

Removal of the instrument signature from and calibration of raw
data are often done by applying standard pipelines offered by the
provider of the facility.  Although they frequently are based upon
familiar legacy systems (e.g. IRAF, MIDAS, CPL/ESO, Starlink, AIPS),
users may find it difficult to disentangle their internal structure,
and be forced to treat them as `black boxes'.  This limits the ability
of a researcher to optimize a pipeline for a particular data set by
either changing its default parameters or by fully replacing some
of its modules.  While a generic pipeline will provide standardized
processing for a wide variety of raw data sets, it will never give
the best results for a specific data set.  Customized processing by
the researcher is often necessary to produce new discoveries.

Reduced, calibrated data are typically obtained through dedicate
pipeline processing or directly from archive data using Virtual
Observatory (VO) tools.  In either case, a set of analysis tools needs
to be applied in order to extract the astronomically interesting
information.  Such tools are still mainly available through legacy
systems although many of them are maintained only at a marginal level
and new developments are currently quite limited.  Most of these legacy
systems were developed to satisfy a certain user community (e.g. with
data originating from special detectors or instruments, often for
wavelength specific regimes).  People analyzing multi-wavelength data
or having special requirements for tools are often forced to use a
combination of several systems as well as stand-alone applications.
Although this is possible, it adds a significant overhead as several,
non-compatible scripting languages and data formats may have to
be applied.  Differences between systems also make it complicated
to compare different application packages for the same purpose and
thereby identify the optimal one for a given set of data.

The current scenario has several disadvantages due to the fragmentation
of efforts.  Each system has to expend effort to follow the general
Information Technology (IT) development such as support of new
operating systems or standards.  This is costly and may lead to legacy
systems being unavailable on modern platforms.  Since the systems
have their own internal standards and conventions, it is difficult
for an individual user to learn them in detail to be able to fully
utilize the facilities provided.  Often the user concludes that it
is simpler to re-write a task even if it is already available in a
system unfamiliar to the user.

\section{Vision for the future}

It is important to recognize the boundary conditions when one considers
how to improve the current situation.  Large organizations with strict
operational requirements as well as large projects with carefully
constrained deliverables may demand full control of code used for
critical tasks \cite{grosbol99}.  On the other side of the scale,
users would like easy access to all available software but without
being forced to use a specific system.  This suggests that any new
software approach must be based upon a modular environment integrating
software from many sources and must depart from the old monolithic,
vertically structured model.  The only solution is to develop
cross-cutting standards to integrate software from different projects
into a common environment.  The primary drivers for establishing a
new common environment \cite{grosbol04} for astronomical software are:

\begin{itemize}
\itemsep=0pt
\item {\bf Efficient access:} A simple, shared interface specification for
  astronomical code would significantly ease sharing and comparison of
  available code and packages.  Applications conforming to such specifications
  could be directly included within and accessed from a common environment.
\item {\bf Separation of astronomy and IT:} A clear distinction between
  astronomical code and IT infrastructure would make it much simpler to
  upgrade or exchange code in the two areas at different timescales.
  It would also make it easier to adopt new IT technology as it becomes
  available and thereby reduce the development and maintenance burden.
\item {\bf Scalability:} With the exponential increase in data volumes
  presently occurring within astronomy, it is becoming increasingly essential
  to develop new algorithms and procedures on an easily accessible, local
  computer and then migrate the software to distributed cluster systems for
  a final processing of large data sets.  Provision of scalability at the
  framework level, making large- and small-scale processing systems comparable
  is critical to making this possible.
\item {\bf Customized implementations:} A single system implementation is seldom
  adequate to encompass the full range of use-cases from small end-user laptops
  to large computational clusters integrating entire observatory data systems.
  A layered, modular concept would allow different organizations to provide
  their own incarnations of the environment, all conforming to the common interface
  specifications, optimized to suit their specific demands while still sharing
  common astronomical code and applications packages.
\end{itemize}

By following these key points, one could establish a body of
astronomical code shared and maintained by the community in the
spirit of open-source, Internet developments. The common interface
specifications would ensure easy access to packages while not forcing
anybody to use a specific implementation of the IT infrastructure.
This would secure the investment in astronomical algorithms even
in a changing IT world.  The needs of observatory data processing
would be addressed while at the same time providing integration with
virtual-observatory standards for distributed multi-wavelength data
analysis capabilities.

\section{Architectural concept}

The OPTICON network on `Future Astronomical Software Environments'
was started in 2004 to discuss requirements, architectural concepts
and high-level design of an environment which would be able to
satisfy future needs of astronomical users for processing and
analysis of their data.  The Network included participants from US
and Europe representing both end-users and people associated to major
data systems.  First, main use-cases were developed such as pipeline
processing, desktop analysis of data, and heavy computational tasks
using clusters of distributed computers.  With these scenarios in mind,
high-level requirements \cite{grosbol05} for future environments were
established and reviewed by the community via the Internet as well
as face-to-face meetings.

\begin{figure}[t]
\begin{center}
\includegraphics[width=0.7\linewidth, angle=0]{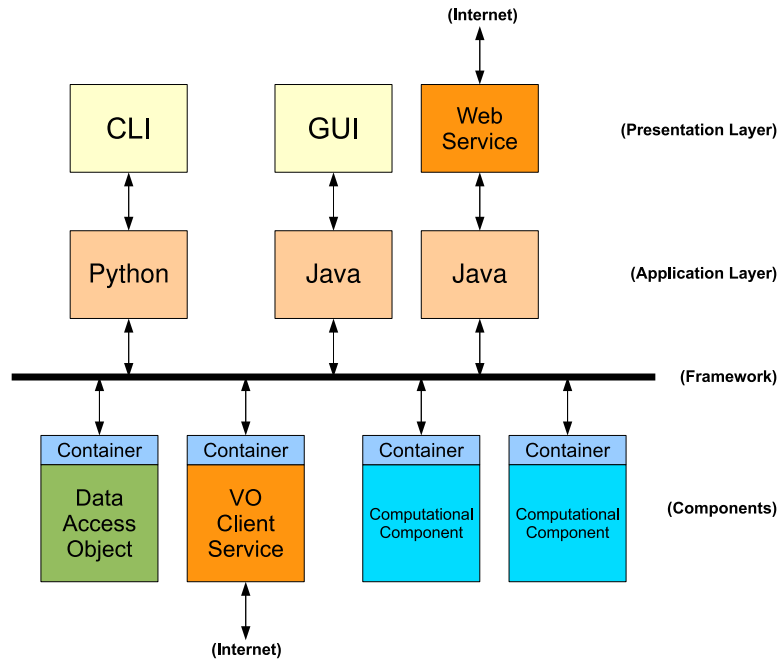}
\caption{Architectural concept for common environment.}\label{fase_fig}
\end{center}
\label{fig1}
\end{figure}

Considering both use-cases and requirements, an architectural
concept \cite{tody06} for future environments was outlined as shown
in Fig.\,1. It is based upon a layered structure (commonly used in the
IT world) and provides support for scalability from laptops to large,
distributed computer systems.  The upper layer (presentation) offers
different options for user interfaces which could range from simple
command line input (CLI) to advanced graphical ones (GUI).  It would
also be possible to provide an interface to other software tools
(e.g. browsers) and thereby establish Web services. The next layer
(application) provides for high-level astronomical applications written
in scripting languages like Python, Ruby or a Unix shell.  In special
cases, one could also use other languages such as Java which offers
easy access to Web, VO, and GUI packages.  Typical applications at
this layer manage computational tasks and the results of such tasks,
but not the computations themselves.  The importance for this layer
is flexibility and easy change rather than optimal performance.

The framework layer connects the application layer scripts with
high performance, computational code at the component layer using
a flexible, distributed communication framework.  This provides a
high level of scalability since computational tasks can be executed
either on a local host or at any available system in the network,
fully transparent to the upper layers.  Asynchronous execution could
also be supported allowing users to perform heavy computational tasks
(e.g. pipeline processing) as background processes.  The framework
would be based upon a {\it software-bus} or {\it messaging} system,
as is commonly used for distributed processing.  By selecting
one of the numerous free implementations (e.g. VO SAMP, D-bus or
openMPI), a major part of the framework would be readily available.
Two other parts would need some astronomical customization, namely:
a) parameter passing mechanism which would transfer parameters and
results between application scripts and computational components,
and b) package management being responsible for the definition and
installation of new astronomical packages into the environment.

The main computational code is placed at the lower layer (organized as
software modules known as `components') which link to the framework
through a standardized interface called the `container'.  On one
side a container connects to the framework while on the other side
it provides a stable, common interface to the astronomical code.
This isolates the astronomical code from the IT infrastructure,
represented by the framework, which often changes faster then
astronomical algorithms.  Separate containers for different coding
languages (e.g. C/C++, FORTRAN, Java) are required but special
binding libraries to ease support for legacy systems are possible.
Different types of components could be envisioned such as simple
computational tasks or stateful services which would allow complex
interaction with the user through GUI's.  Special components to access
data or Web services (e.g. provided by VO) could also be made.

This architectural concept has many advantages compared with monolithic
systems and independent developments.  The clear separation between
astronomical code and IT infrastructure makes it easy to obtain
the latter through public-domain IT projects and thereby reduce
maintenance cost.  Further, the astronomical parts can be migrated to
new IT infrastructures with a minimum of labor.  Organizations that
demand full control can implement their own environment but still
benefit from algorithms using the common interface specifications.

A common, accepted interface to the environment would simplify
the integration of new astronomical code.  Individual astronomical
developers would have direct access to the environment and its general
facilities by using the standardized interface.  Their code would also
be readily available to the community through a simple packaging and
package distribution mechanism.

\section{Way forward} With general agreement on this architectural
concept \cite{tody07}, the OPTICON network with the USVAO have started
a detailed design study to outline the main parts of the environment
and their interfaces.  The next step is a prototype implementation of
the most important parts (i.e. parameter passing, package management,
messaging and execution) as a proof-of-concept.  This will lead to
a refinement of interface specifications and structure.  Finally, a
minimal reference implementation for a desktop system is foreseen to
provide a test-bed which can be evaluated by the community. It would
be available initially for at least the Linux and MacOSX operating
systems with support for Python for scripting, Java for VO and Web
access, and C/C++/FORTRAN for compiled code.  Parallel to this effort,
current legacy systems will be contacted to evaluate the feasibility
of adapting their most important packages to the new environment.

Making a basic version of the environment available to astronomers will
not convince them that it is worthwhile using it.  This can only be
done by making it easy to install and handle on common desktop systems.
The version must provide access to important legacy applications and
allow new software to be added in a trivial manner.

The long-term success of the concept proposed will depend on other,
more administrative factors.  The common interface specifications must
be stable which require them to be controlled by an international
body (e.g. an IAU working group as for the FITS standard, or an
appropriate subgroup operating within the IAU or IVOA frameworks).
Basic support and maintenance of a reference environment have to be
secured even if it only provided a simple desktop version.

\section{Conclusions}

The current situation with a multitude of largely incompatible legacy
or more modern but narrowly focused systems for data processing and
analysis is not satisfactory since it makes access to software packages
more difficult and requires significant resources for maintenance.
To remedy these problems, a new approach is required based upon the
concept of a common astronomical software framework which allows
flexible integration of both new and legacy code, which is scalable
to leverage modern computational hardware architectures while meeting
the challenge of exponentially increasing data volumes, and which
separates astronomical code from the IT infrastructure to provide
stability for critical  astronomical code while allowing rapid uptake
of new IT technologies.

The requirements and architectural concept for such an environment have
been discussed within an OPTICON Network with US and EU participation.
It is proposed to adopt a modular, layered model based upon a
framework supporting messaging and execution in a distributed system.
Major parts of the framework can be taken directly from general
open-source IT projects while other areas, like parameter passing,
need to be customized for astronomical applications.  The isolation
of IT infrastructure will safeguard the astronomical code against the
rapid changes in the IT world.  The generic, modular structure would
allow easy integration of both new and legacy packages.  This would
significantly increase the efficiency with which astronomers can
access a wide variety of available astronomical software.  An efficient
access to software tools would directly lead to accelerating the rate
of astronomical discovery.

\section*{Acknowledgements}
This paper is based on discussions in the OPTICON Network 3.6 on 'Future
Astronomical Software Environments'.  We especially want to thank INAF for its
support of this work.  The project has been supported by OPTICON funded by
the European Commission under Contract no. {\tt RII3-CT-2004-00001566} and the
USNVO under NSF Cooperative Agreement {\tt AST0122449}.

\end{document}